\documentclass[10pt]{article}
\usepackage[margin=1.3in]{geometry}
\usepackage[utf8]{inputenc}
\usepackage[]{graphicx}
\usepackage{amsmath,amssymb,amsfonts}
\usepackage{lipsum}
\usepackage[compress]{cite}
\def\email#1{\it #1}

\def\keywords#1{\par
	\vspace*{8pt}
	{\leftskip18pt\rightskip\leftskip
			\noindent{\it Keywords}:\ #1\par}\par}

\begin{document}

\title{\textsc{On slowly rotating  magnetized white dwarfs}}


\author{Diana Alvear Terrero${}^{1,\ast}$, Daryel Manreza Paret${}^{2,3, \dagger}$,\\ Aurora Perez Martinez${}^{1,3, \ddagger}$\\[10pt]
	${}^1\,$Departamento de Física Teórica,  Instituto de Cibern\'{e}tica Matem\'{a}tica y F\'{\i}sica  \\
 	Calle E esq 15 No. 309, Vedado, La Habana 10400, Cuba
 	 \\[5pt]
	${}^2\,$Facultad de F{\'i}sica,  Universidad de la Habana
	\\
	 San L{\'a}zaro y L, Vedado, La Habana 10400, Cuba 
	 \\[5pt]
	${}^3\,$Instituto de Ciencias Nucleares,  Universidad Nacional Autónoma de México
	\\
	A. P.70-543, 04510 C. México, México 
	\\[8pt]
	${}^{\ast}$\email{dianaalvear@icimaf.cu},
	${}^{\dagger}$\email{dmanreza@fisica.uh.cu},
	${}^{\ddagger}$\email{aurora@icimaf.cu}
}

\maketitle

\begin{abstract}
Rotating magnetized white dwarfs are studied within the framework of general relativity using Hartle's formalism. Matter inside magnetized white dwarfs is described by an equation of state of particles under the action of a constant magnetic field which introduces anisotropic pressures. Our study is done for values of magnetic field below $10^{13}$ G -a threshold of the maximum magnetic field obtained by the cylindrical metric solution- and typical densities of WDs. The effects of the rotation and magnetic field combined are discussed, we compute relevant magnitudes such as the moment of inertia,  quadrupole moment and eccentricity.
\end{abstract}


\section{Introduction}	

White dwarfs are very well-known compact objects with typical values of mass around a solar mass and the size of the Earth. Composed mainly by carbon, they counteract the gravitational pull by means of the pressure of the degenerate electron gas while the carbon nuclei are the principal contribution to the mass.

Observations estimate magnetized white dwarfs (MWDs) surface magnetic fields in the range of  $10^6\,$G to $10^9\,$G\cite{Terada:2007br,Schmidt:1995eh} whereas internal magnetic fields are determined indirectly using theoretical models based  on macroscopic and microscopic analyses. Moreover,  there are observations of superluminous thermonuclear supernovae, whose progenitor could be
super-Chandrasekhar WDs \mbox{($M_{\textrm{\scriptsize{WD}}}>1.44M_{\odot}$)} \cite{Howell:2006vn}. Consequently, in Refs.~\cite{Das:2012ai,Das:2013gd,Das:2014ssa} it was proposed to justify their existence with the presence of strong magnetic fields above $10^{13}\,$G.

A magnetic field acting on a fermions system breaks the SO(3) symmetry, giving rise to an anisotropy in the equations of state (EoS) \cite{2000PhRvL..84.5261C}.

Furthermore, the anisotropy of the energy momentum tensor caused by the magnetic field can be included considering an axi-symmetric and poloidal strong magnetic field\cite{PhysRevD.92.083006}, which allows to model rotating magnetized white dwarfs in a self-consistent way by solving Einstein-Maxwell equations\cite{Bonazzola:1993zz,Bocquet:1995je}.

The presence of the anisotropic pressures suggests that introducing an axially symmetric metric to solve Einstein equation is crucial. Previously we have used a cylindrical metric and obtained a maximum bound of $1.5\!\times\!10^{13}$ G for the magnetic field of stable MWDs \cite{1674-4527-15-10-1735}, ruling out the possibility of super-Chandrasekhar WDs with strong magnetic fields. 

Rotation is another plausible cause for the increment in mass of white dwarfs.
In order to investigate this issue, we solve the rotating structure equations emerging from spherical symmetry by Hartle's method --despite spherical metric is no longer adequate when considering anisotropic EoS--. This allows us to determine if the deformation of the rotating stars accounts for stable RMWDs with mass above $1.44M_{\odot}$.

With that aim, we first describe the equilibrium of RMWDs by solving Einstein equations in section \ref{sec:srecs} while considering both pressures -one parallel and the other perpendicular to the magnetic field- independently. Then, in section \ref{sec:res} we present numerical results, and finally in section \ref{sec:concl}, our conclusions.

\section{Slowly rotating structure equations for RMWDs} \label{sec:srecs}

When discussing the structure of compacts objects, it must be analyzed both local and global properties of the involved matter. The first ones are described by an equation of state (EoS), while the latter ones comprises the dynamical response of matter at large scales to, for instance, gravity and rotation. In this paper, we consider the magnetized equations of state obtained in Refs.~\cite{ASNA:ASNA201512236,2016arXiv160100832A}  for carbon/oxygen WDs whose matter is composed by particles under the action of a constant magnetic field, which leads to a splitting of the pressure into a component parallel to the magnetic field and a perpendicular one. The values of the magnetic field are chosen below the $10^{13}$ G threshold mentioned before.

 Regarding the structure equations for a slowly rotating compact object, we take into account the angular velocity ($\Omega$) of the star  uniform and sufficiently slow so that $R^3\Omega^2 \ll M$, where M and R are the mass and the radius of the non-rotating WDs respectively.  Then, the angular velocity  provokes small changes in the pressure $P$, energy density $E = \rho c^2$ and gravitational field with respect to the corresponding quantities of the static configuration. These changes can be considered as perturbations of the non-rotating solution.

So,  to consider that a star is rigidly and slowly rotating implies calculating its equilibrium properties reckoning small perturbations on static configuration.  Introducing new coordinates $(r,\theta,\phi)$, where $r(R,\theta) = R + \xi (R, \theta)$ takes into account deviations from spherical symmetry, the metric of the rotating configuration becomes \cite{1967ApJ...150.1005H,1968ApJ...153..807H}

\begin{eqnarray} \nonumber
ds^2 & = & -\, e^{\nu}\left\{1+2\left[h_0 + h_2 P_2(\cos \theta)\right]\right\}  dt^2
+ \frac{1+2\left[m_0+m_2 P_2(\cos \theta) \right]\left[r-2M \right]^{-1}}{1- 2M/r} \, dr^2 \\[1pt]
&& +\, r^2\left[1+2(v_2-h_2)P_2(\cos \theta)\right] \left[d\theta^2 + \sin^2\theta \left(d\phi - \omega dt\right)^2\right] + O(\Omega^3).
\label{eq:metric}
\end{eqnarray}

Here $P_2(\cos \theta)$ is the Legendre polynomial of second order, $e^{\nu}$ and \mbox{$e^{\lambda} = \left[1 - 2M(r)/r\right]^{-1}$} are the static metric functions, and \mbox{$\omega (r) = \bar \omega (r) + \Omega$} is the angular velocity of the local inertial frame, where $\bar \omega (r)$ is the fluid's angular velocity relative to the local inertial frame. The functions $h_0\!=\! h_0(r)$, $h_2\!=\! h_2(r)$, $v_2\!=\! v_2(r)$, and mass perturbation factors $m_0\!=\! m_0(r)$ and $m_2\!=\! m_2(r)$ are all proportional to $\Omega^2$. Besides, we must define the pressure perturbation factors $p_0^\ast$ and $p_2^\ast$ on the order of $\Omega^2$, which modify the energy-momentum tensor \cite{1968ApJ...153..807H}.

Once computed Einstein equations considering perturbations up to $O(\Omega^2)$ with the metric (\ref{eq:metric}), the structure of the perturbed rotating stars is described by the static equations of Tolman-Oppenheimer-Volkoff for the pressure $P$, the mass and $\nu$ in addition to the equations for $\bar{\omega}$, $m_0$, $p_0^\ast$,  $h_2$, $v_2$, $m_2$ and $p_2^\ast$. The system to integrate outward is

	\begin{eqnarray}
	\frac{dP}{dr} &=& -\frac{(E+P)(M+4\pi r^3 P)}{r(r-2M)} , \label{TOV1} \\[1pt]
	\frac{dM}{dr} &=& 4\pi r^2 E  , \label{TOV2}  \\[1pt]
	\frac{d\nu}{dr} &=& -\frac{2}{E+P}\frac{dP}{dr} , \label{TOV3} \\[1pt]
    \frac{d\bar{\omega}}{dr} & =& \kappa, \\[1pt]
	\frac{d \kappa }{dr}& =&  \frac{4\pi r (E+P)(r \kappa+4\,\bar \omega)}{r-2M} - 4 \frac{\kappa}{r} , \\[1pt]
\frac{dm_0}{dr} & = & 4\pi r^2 (E+P)\frac{dE}{dP} p_0^\ast+ \frac{r^3e^{-\nu}}{3}\left(r-2M\right) \left[ \frac{\kappa^2}{4} + \frac{8\pi r(E+P) \,\bar\omega^2 }{r-2M}\right],  \\[1pt]
\frac{dp_0^\ast}{dr}& =& \!-\frac{m_0(1+8\pi r^2P)}{(r-2M)^2} \!-\! \frac{4\pi r^2 (E+P)}{r-2M}p_0^\ast \!+  \!\frac{r^3e^{-\nu}}{3} \left[ \frac{\kappa^2}{4} +\!  \frac{\bar \omega^2}{r}\! \left(\!\frac{2}{r}-\frac{d\nu}{dr}\!\right)\!+\!\frac{2\,\kappa\, \bar \omega}{r}\!\right]\!, \;\\[1pt]
\frac{dv_2}{dr} & = & -h_2 \frac{d\nu}{dr}+\frac{r^3e^{-\nu}}{3}\left(r-2M\right) \left[\frac{2}{r}+\frac{d\nu}{dr}\right]\left[ \frac{\kappa^2}{4} + \frac{4\pi r(E+P) \,\bar\omega^2 }{r-2M}\right],  \\[1pt] \nonumber
\frac{dh_2}{dr} & = & h_2 \! \left[-\frac{d\nu}{dr} \!+ \! \frac{8\pi r^3(E+P)-4M}{r^2(r-2M)}\left(\!\frac{d\nu}{dr}\!\right)^{\!\!-1} \right]  -  \frac{4 v_2}{r(r-2M)}\left(\!\frac{d\nu}{dr}\!\right)^{\!\!-1} \\[1pt] \nonumber
						&& + \, \frac{r^3 e^{-\nu}}{3} \left(\!\frac{d\nu}{dr}\!\right)^{\!\!-1}\! \left\{
						\frac{\kappa^2}{4} \left[\left(r-2M\right)\left(\!\frac{d\nu}{dr}\!\right)^{\!2}-\frac{2}{r}\right] \right.\\[1pt]
						&& \, \left. + \,\frac{4\pi r(E+P)\,\bar\omega^2}{r-2M} \left[(r-2M)\left(\!\frac{d\nu}{dr}\!\right)^{\!2} +\frac{2}{r}\right] \right\},
\end{eqnarray}
alongside with expressions
\begin{eqnarray}
	m_2 &=& (r-2 M) \left\{\frac{1}{6}r^3e^{-\nu}\left[(r-2M)\,\kappa^2 + 16 \pi r (E+P)\, \bar \omega^2 \right]-h_2\right\} , \\
	p_2^\ast &=& - \frac{1}{3}r^2e^{-\nu} - h_2\,.
\end{eqnarray}

These equations must be solved with the proper boundary conditions. This means to contemplate values of the central energy density within typical values for WDs. The pressure is maximum  in the center of the star  and must go to zero at the surface. Hence, the integration is carried out until $P$ vanishes. The value for $\bar  \omega$ at the center is arbitrary 
and the rest of the variables are set up to zero initially.

The total angular momentum is $J=R^4\kappa(R)/6$, the angular velocity of the rotating WD is \mbox{$\Omega = \bar \omega (R) + 2J/R^{3}$}, the moment of inertia is $I =J/\Omega$, and the total mass is

\begin{small}
\begin{equation}
	M_T = M(R) +	m_0(R)	+ \frac{J^2}{R^3}\,.
\end{equation}
\end{small}

Also, we compute the quadrupolar momentum \cite{1967ApJ...150.1005H,1968ApJ...153..807H}

\begin{small}
	\begin{equation}
	Q = \frac{8}{5} M^3 \frac{h_2+v_2-\tfrac{J^2}{M R^{3^{\phantom 1}\!\!}}}{ \frac{2M}{\sqrt{R(R-2M)}^{\phantom A} } Q_2^{{\phantom 2}1} \left(\tfrac{R}{M} -1\right) + Q_2^{{\phantom 2}2}\left(\tfrac{R}{M} -1\right)} + \frac{J^2}{M},
	\end{equation}
\end{small}

\noindent where $Q_m^{{\phantom m}n}(x)$ is the associated Legendre function of second kind. The rotational deformation of the WD can be depicted through the displacement of the surface of constant density at radius $r$ in the static configuration to

\begin{small}
\begin{eqnarray}
 &&	r(R,\theta) = R + \xi_0 (R) + \xi_2 (R) P_2(\cos \theta), \\[1pt]
 &&	\xi_0 =  - p_0^\ast (E+P) \! \left(\frac{dP}{dr}\right)^{\!-1}\!,   \quad
	\xi_2 = - p_2^\ast (E+P) \! \left(\frac{dP}{dr}\right)^{\!-1}\!
	\end{eqnarray}
\end{small}
when rotating. The eccentricity is
\begin{small}
	\begin{equation}
	\varepsilon = \sqrt{1-\left(\frac{R_p}{R_eq}\right)^2},
	\end{equation}
\end{small}
with
	\begin{eqnarray}
		R_p &=&  r (R, 0)  = R + \xi_0 (R) + \xi_2 (R),  	  \\[1pt]
		R_{eq} &=& r(R,\tfrac{\pi}{2})=R + \xi_0 (R) - \frac{\xi_2 (R)}{2}.
	\end{eqnarray}

\section{Results and discussion} \label{sec:res}

\begin{figure}[h] 
	\centering
		\includegraphics[width=0.5\linewidth]{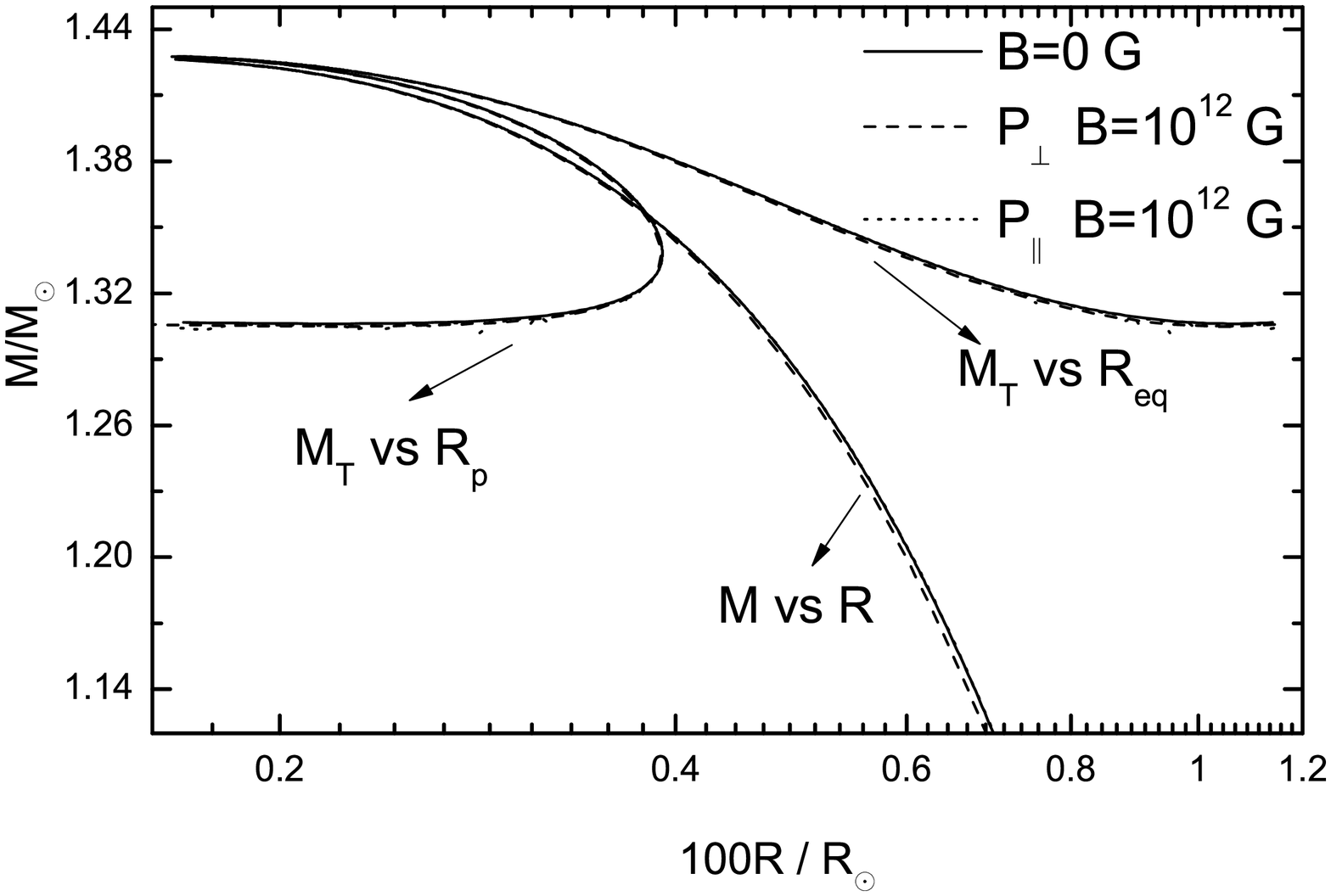}
		\includegraphics[width=0.49\linewidth]{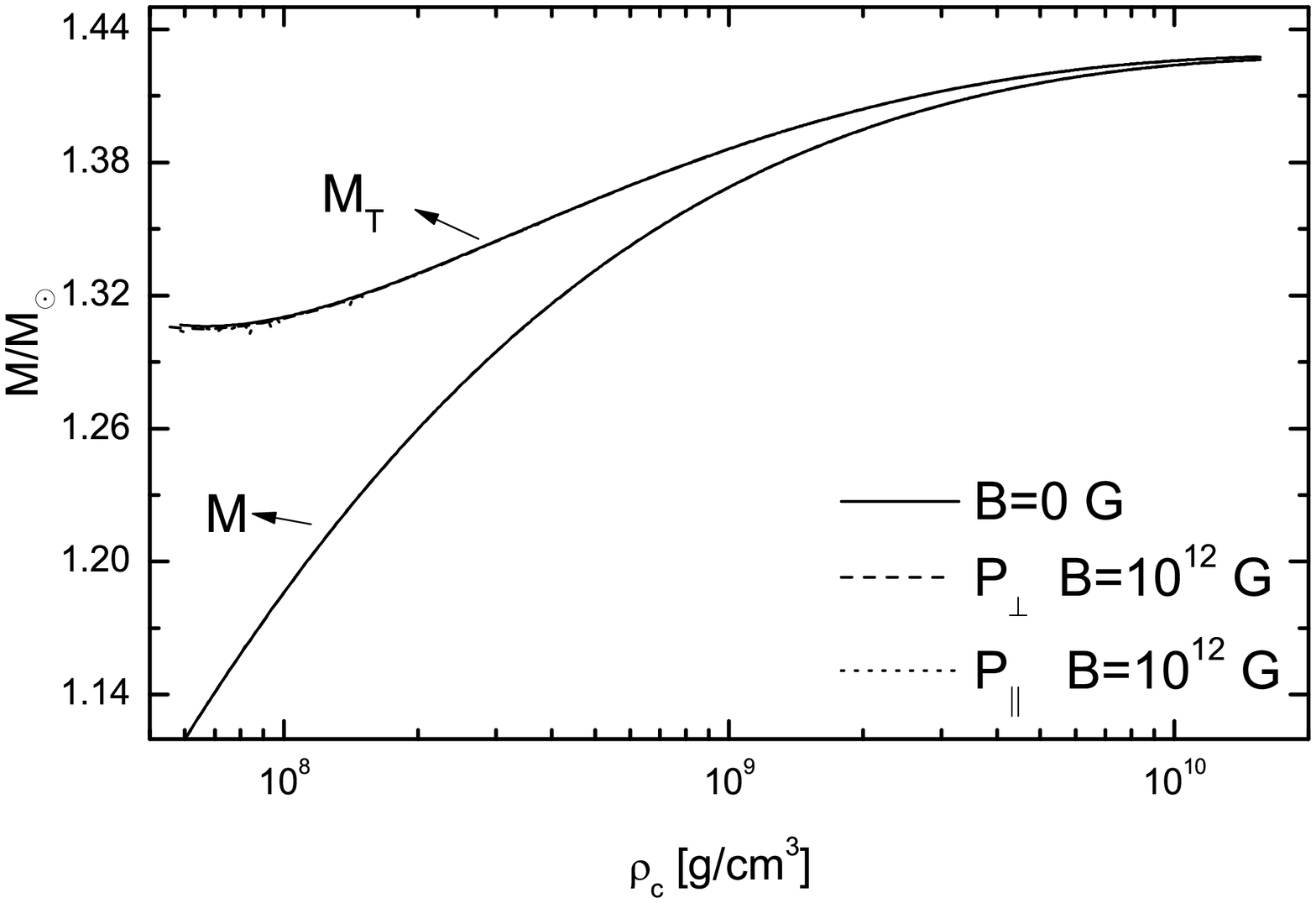}
	\caption{Left panel: Mass versus radius for static configuration and total mass as a function of equatorial and polar radii. Rigth panel: Static and total masses as a function of central density.}\label{fig:MRrho}
\end{figure}

Fig.~\ref{fig:MRrho} shows the behavior of the mass for the static and the rotating configurations. In the left panel, we superpose the curves corresponding to $M$ versus $R$, $M_T$ versus $R_p$  and $M_T$ versus $R_{eq}$. In the right panel, both masses are shown as a function of the central density. All plots include the non-magnetic configuration as well as the solutions for the parallel and perpendicular pressures corresponding to $B=10^{12}$ G. 

As described in the previews section, considering slow rotation increases the mass of the stars. However, this increment diminishes as the density increases, so that the outcome for the total mass is lower than the Chandrasekhar mass even for higher densities solutions, at least for the values of the magnetic fields below the Schwinger critical magnetic field for which our EoS are valid.  Furthermore, the precision of our stable solutions increases with the central density of the star.

\begin{figure}[ht] 
	\centering
		\includegraphics[width=0.5\linewidth]{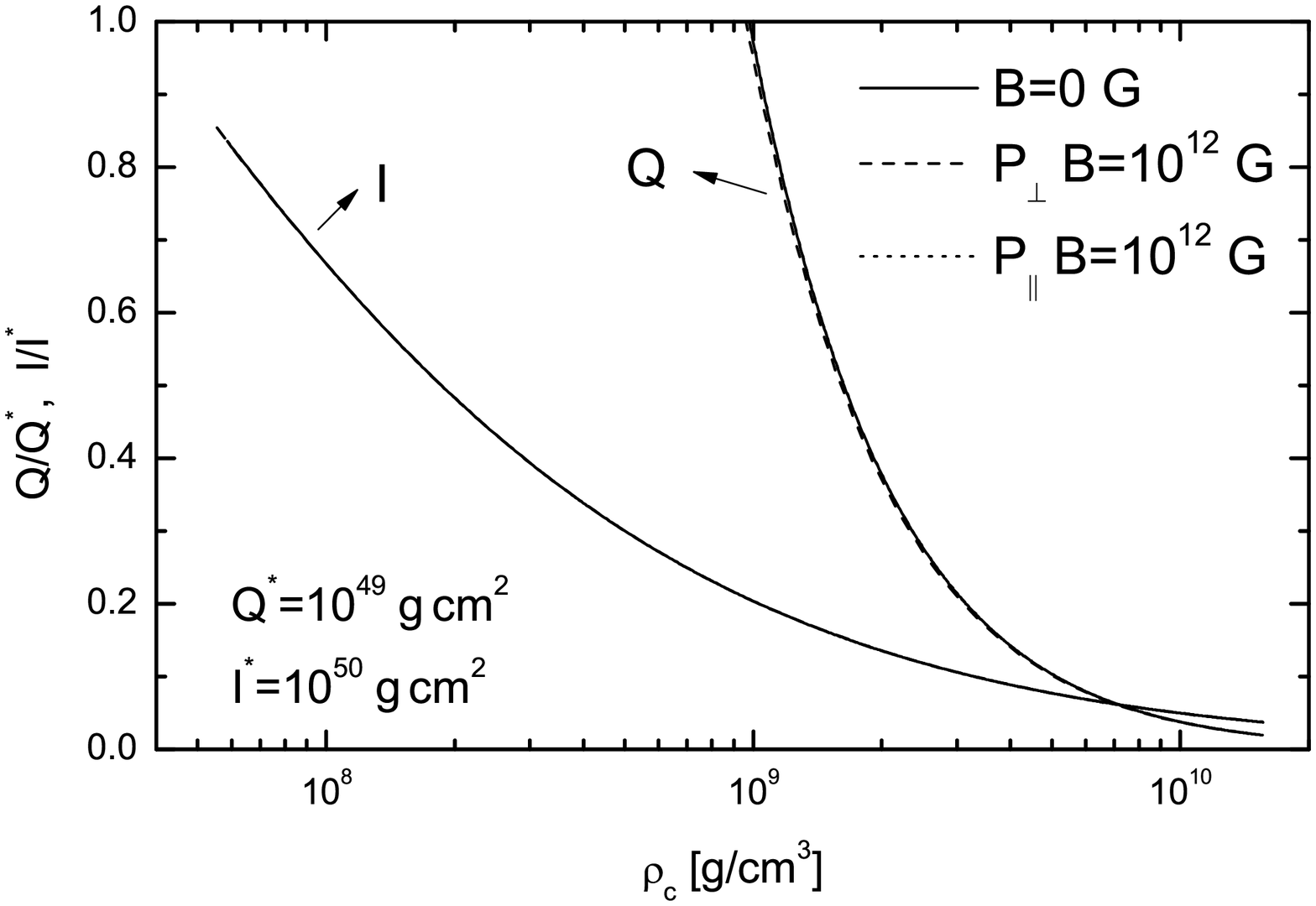}
		\includegraphics[width=0.49\linewidth]{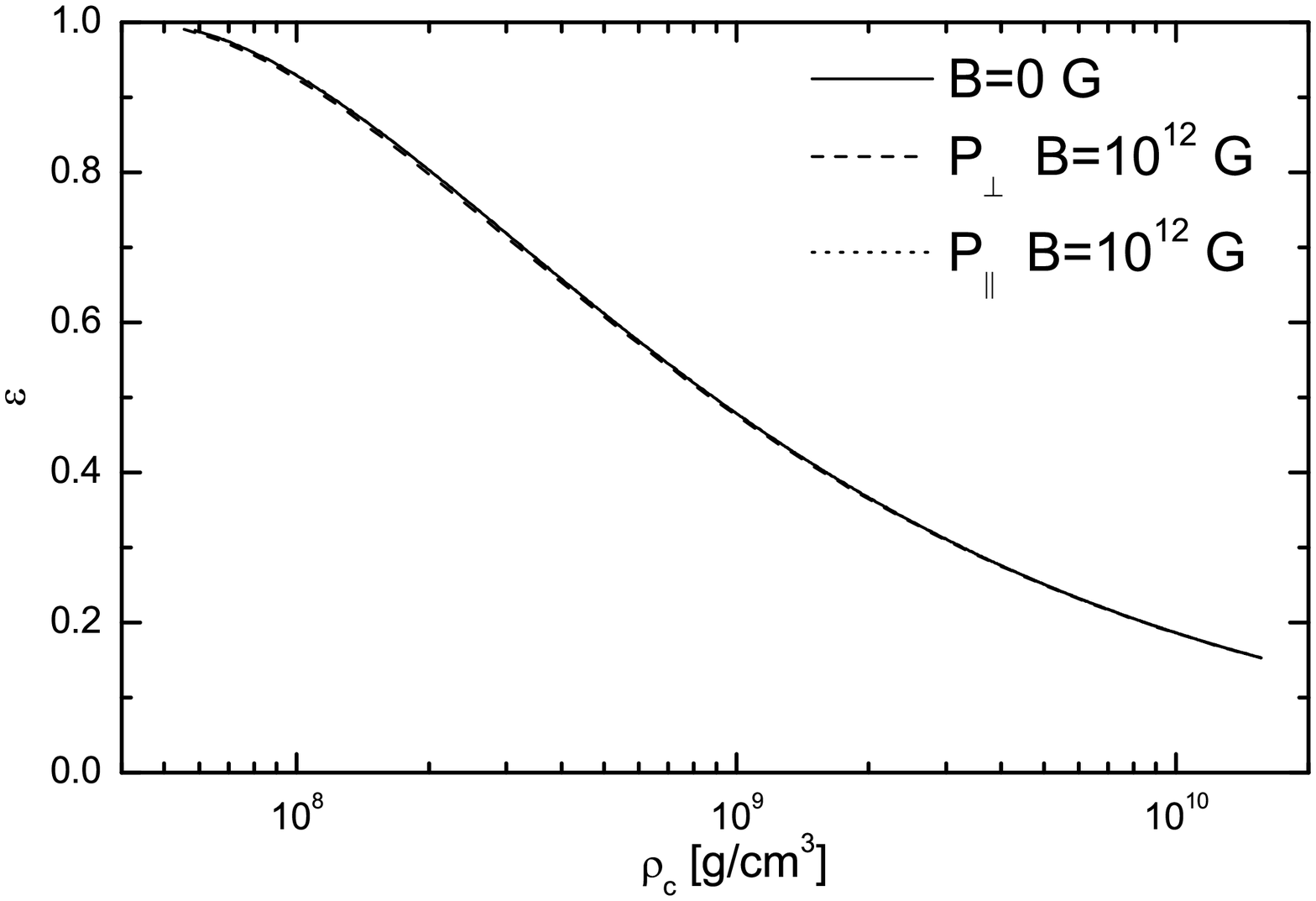} 
	\caption{Left panel: Moment of inertia and quadrupolar momentum as a function of central density. Right panel: Eccentricity versus central density.}
	\label{fig:IQEx}
\end{figure}

In left panel of Fig.~\ref{fig:IQEx} we present the moment of inertia $I$ and the quadrupolar momentum $Q$ as a function of density for $B=0$, and $B= 10^{12}$ G, for both parallel and perpendicular pressures. The right panel shows the eccentricity also as a function of $\rho_c$.  The change of the magnetized solutions respect to the non-magnetic ones are not substantial. Contrary to the behavior of $I$ and $Q$, the eccentricity decreases with the increment of energy density. 

\begin{figure}[htb] 
	\centering
	\includegraphics[width=0.51\linewidth]{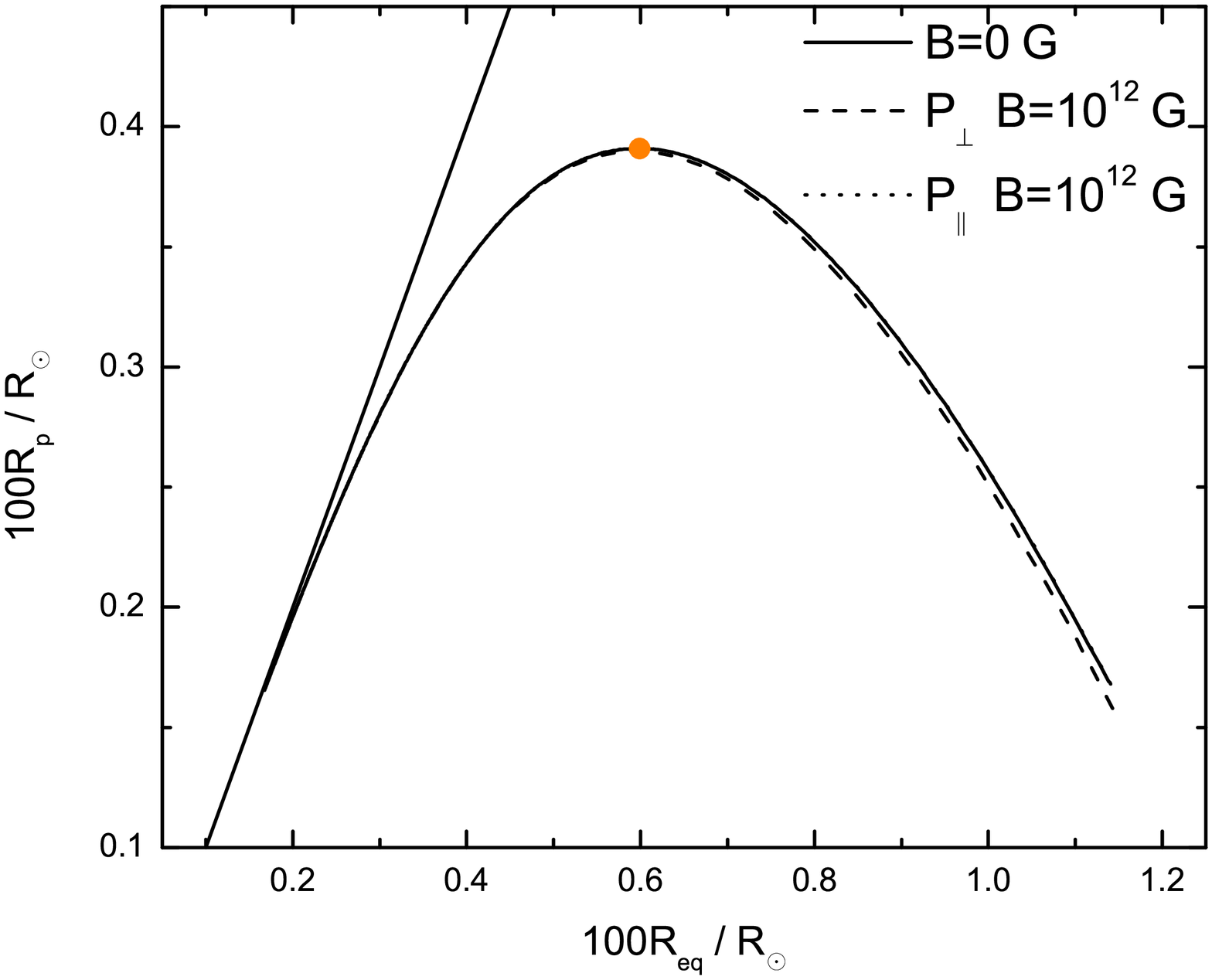}\hspace*{0.01\linewidth}
	\includegraphics[width=0.4\linewidth]{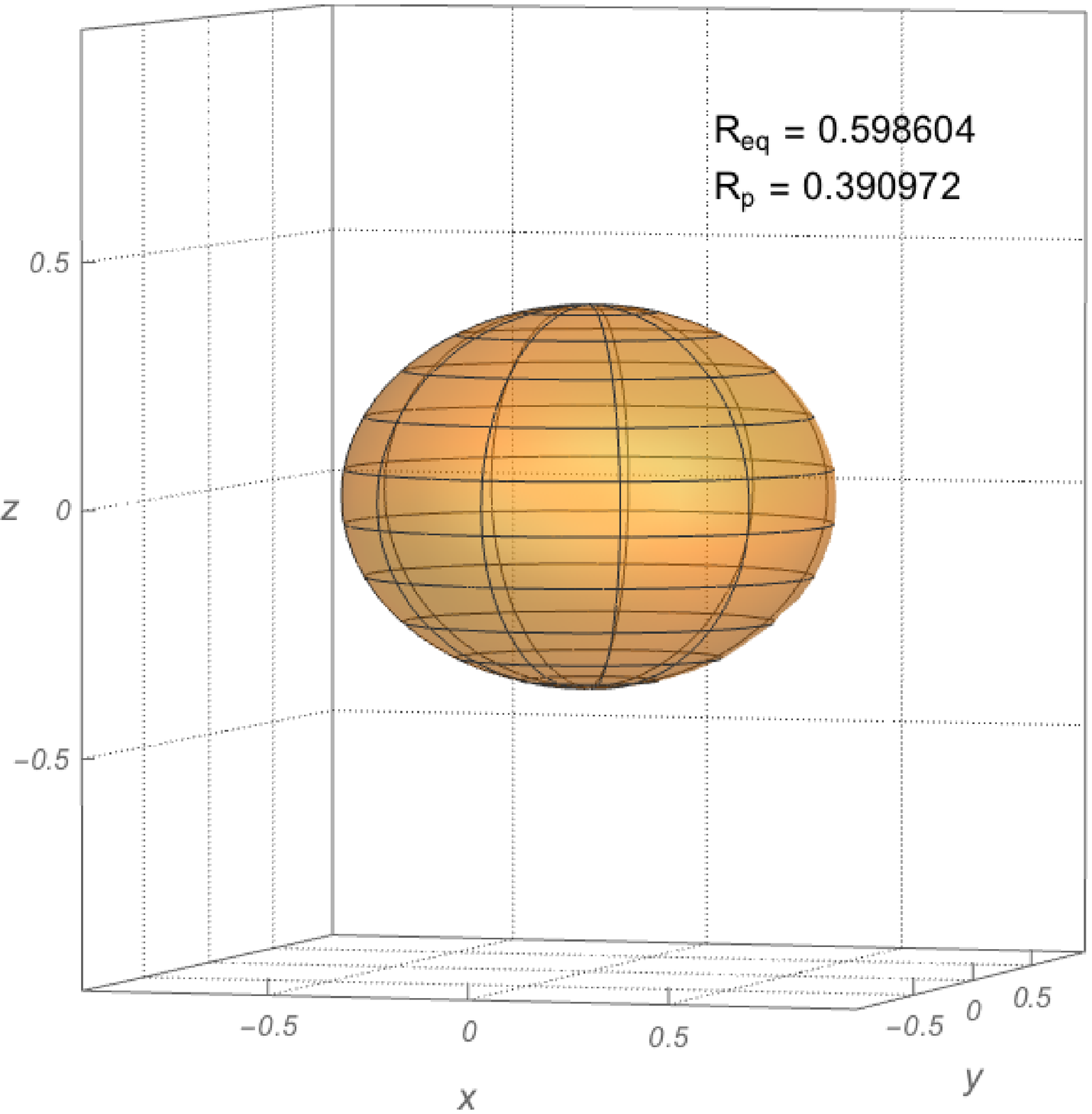}
	\caption{Left panel: Polar radius versus equatorial radius, with density increasing towards left. The orange point corresponds to the $B=0$ solution at $\rho_c =2.49547\times 10^8$ g cm$^{-3}$. Right panel: representation of the oblate spheroid shape of the WD associated to the orange point in the graph of the left panel.}
	\label{fig:Sph}
\end{figure}

Additionally, the facts that $0\!<\!\varepsilon\!<\!1$, $R_p\! <\! R_{eq}$ and $Q\!>\!0$ implies that the solutions correspond to oblate WDs configurations. This can be easily pictured from Fig.~\ref{fig:Sph}, where we have plotted the polar radius versus the equatorial radius on left panel, and, in the right panel we have constructed a parametrical surface of the non-magnetized solution at $\rho_c =2.49547\times 10^8$ g cm$^{-3}$ using the corresponding values of $R_{eq}$ and $R_p$. 

\section{Conclusions} \label{sec:concl}

We have implemented an algorithm to study slowly rotating MWDs using the formalism proposed by Hartle for magnetized equations of state. Numerical solutions have been computed for the total mass of the rotating star as well as its equatorial and polar radii and the static couterpart. The moment of inertia, the quadrupolar momentum and the eccentricity were analyzed, confirming that rotation in this way deforms the star, that are now oblate spheroids. In all cases, results were obtained for the non-magnetic configuration and for a fixed value of $10^{12}$ G, lower than $10^{13}$ G, a critical field beyond which solutions are unstable. Our results for non-magnetized slowly rotating WDs are in agreement with Refs.~\cite{Boshkayev:2012bq,doi:10.1093/mnras/stw2614}. Also, the stable slowly rotating solutions obtained are bounded by the condition of applicability of Hartle's method, which is satisfied more accurately for WDs of higher densities.

Taking into account the splitting of the pressure, it would be interesting to investigate the possibility of an alternative method to the one discussed here in order to include both, parallel and perpendicular pressures at the same time. This could give an insight into a more precise description of such anisotropic WDs, and would allow to compare the deformation of the stars due to the effect of the magnetic field with the rotational deformation.
 
\section*{Acknowledgements}

D.A.T, D.M.P and A.P.M  have been supported by the grant CB0407 and the ICTP Office of External Activities through NET-35. A.P.M thanks Consejo Nacional de Ciencia y tecnologia (CONACYT) for the support with the sabbatical Grant 264150 at ICN-UNAM, M\'exico, where this work was developed.  D.M.P  has been also supported by a DGAPA-UNAM fellowship.

\end{document}